\documentclass[prd,reprint,
 amsmath,amssymb,
 aps,
superscriptaddress,
nofootinbib]{revtex4-1}

\usepackage{placeins}
\usepackage{graphicx}
\usepackage{dcolumn}
\usepackage{bm}
\usepackage[hidelinks,colorlinks=true,allcolors=blue]{hyperref} 
\usepackage[capitalise]{cleveref} 
\usepackage{enumerate}
\usepackage{subfigure}

\begin{document}

\preprint{APS/123-QED}

\title{Early Dark Energy During Big Bang Nucleosynthesis}

\author{David McKeen}
\email{mckeen@triumf.ca}
\affiliation{TRIUMF, 4004 Wesbrook Mall, Vancouver, BC V6T 2A3, Canada}

\author{Afif Omar}
\email{aomar@triumf.ca}
\affiliation{TRIUMF, 4004 Wesbrook Mall, Vancouver, BC V6T 2A3, Canada}
\affiliation{Department of Physics and Astronomy, University of Victoria, Victoria, BC V8P 5C2, Canada}

\date{\today}

\begin{abstract}
We study the impact of an early dark energy component (EDE) present during big bang nucleosynthesis (BBN) on the elemental abundances of deuterium D/H, and helium $Y_p$, as well as the effective relativistic degrees of freedom $N_{\rm eff}$. We consider a simple model of EDE that is constant up to a critical time. After this critical time, the EDE transitions into either a radiation component that interacts with the electromagnetic plasma, a dark radiation component that is uncoupled from the plasma, or kination that is also uncoupled. We use measured values of the abundances and $N_{\rm eff}$ as determined by cosmic microwave background observations to establish limits on the input parameters of this EDE model. In addition, we explore how those parameters are correlated with BBN inputs; the baryon-to-photon ratio $\eta_b$, neutron lifetime $\tau_n$, and number of neutrinos $N_\nu$. Finally, we study whether this setup can alleviate the tension introduced by recent measurements of the primordial helium abundance.
\end{abstract}

\maketitle


\section{\label{sec:level1}Introduction}
The earliest direct probe of the Universe's history is the formation of light elements when it was a few minutes old, during a process called big bang nucleosynthesis (BBN). The successful predictions of light element abundances, particularly $^2{\rm H}$ and $^4{\rm He}$, in a standard cosmological history places strong bounds on physics beyond the standard model (SM). 

Big bang nucleosynthesis takes place as the Universe cools from a temperature of around a few MeV to tens of keV. In the standard $\Lambda$CDM cosmological history, at this time most of the energy of the Universe is stored in relativistic species in the form of photons, three flavors of neutrinos, and electrons and positrons until they become nonrelativistic and annihilate. Because of the relatively low density of baryons, nucleosynthesis proceeds via two-body reactions, with the main impediment being the formation of deuterium due to its low binding energy---this deuterium bottleneck is not cleared until temperatures of order $T_{\rm bn}\equiv 70~\rm keV$, an order of magnitude below its binding energy because of the large photon-to-baryon ratio. Once deuterium is produced heavier nuclei such as tritium and helium can be formed.

Since, in the standard cosmological picture, the energy budget of the Universe is ``radiation dominated'' during this epoch, the temperature evolves in time as $T\propto t^{-1/2}$. The successful predictions of BBN rely on this relationship which governs the opening of the deuterium bottleneck and the subsequent formation of heavier nuclei, which is determined by complicated nuclear physics cross sections that depend on the temperature of the Universe.

Periods where vacuum energy played an important role have been motivated both observationally and theoretically. As is well known, observations of the accelerating expansion rate of the Universe today tell us that the current energy budget is predominantly in such a (different) component today which behaves like a cosmological constant. The cosmological horizon and flatness problems are alleviated if the expansion during earliest moments after the big bang was driven by another form of dark energy, a process called inflation, that transformed into a different form during ``reheating''. Recently, the possibility that a non-negligible fraction of the Universe's energy density transformed as ``early dark energy'' (EDE) around the time of recombination has been studied as a potential solution to the tension between early- and late-time measurements of the expansion rate of the Universe (see, e.g.~\cite{Karwal:2016vyq,Mortsell:2018mfj,Poulin:2018cxd,Smith:2020rxx,Schoneberg:2021qvd,Vagnozzi:2023nrq,Sobotka:2023bzr}). The enhanced absorption of 21-cm radiation reported by the EDGES Collaboration~\cite{Bowman:2018yin} can be explained by a period of EDE, although not without increasing tension with cosmic microwave background (CMB) observations~\cite{Hill:2018lfx}. More speculatively, the zero-point energy density in the cores of neutron stars could differ substantially from that in the vacuum due to a phase transition in hadronic matter, potentially leading to observable effects in mergers~\cite{Csaki:2018fls}.

Big bang nucleosynthesis~\cite{Schramm:1977ne,Bernstein:1988ad,Walker:1991ap,Cyburt:2015mya} has long been a test of cosmology and particle physics~\cite{Sarkar:1995dd,Iocco:2008va,Jedamzik:2009uy,Pospelov:2010hj}. In Ref.~\cite{Kaplinghat:2000zt}, the BBN effects of ``power law cosmologies'' with differing expansion histories were studied, showing that the standard radiation-dominated picture was favored by data on light element abundances. In this paper, we explore the impact of variations of this picture of radiation domination during BBN. In particular, we study the effect of EDE that briefly becomes significant when compared to radiation energy density during BBN. This could be achieved by considering dynamic models of dark energy as a scalar field like quintessence or perhaps a scalar field sitting in a potential that undergoes a phase transition (e.g.~\cite{Caldwell:1997ii,Doran:2006kp,Gomez-Valent:2022bku,Wetterich:2004pv,Poulin:2018cxd} and see~\cite{Lee:2022cyh} for details about the interplay between EDE and scalar fields) . We parameterize the EDE by the energy density during its dark energy phase as well as the time at which it transitions from dark energy to a new form. We solve the equations that govern the production of light elements and compare to observations, constraining parts of this parameter space. We also consider recent measurement of the primordial $^4{\rm He}$ abundance that differs from what is expected in standard BBN, and show that this tension could be somewhat alleviated with EDE.

The paper is outlined as follows: In \cref{cosmological eq}, we present the equations governing
$\Lambda$CDM cosmology and some of the simplifying assumptions taken. In \cref{Tt}, we write down the system of equations that describes the thermal evolution of the Universe necessary for solving the Boltzmann equations for the BBN nuclear reaction network. In \cref{SectionBBN}, we detail how the nuclear reaction network is solved to produce standard BBN. In \cref{EDE_BBN} we introduce EDE into the model and explore three different scenarios for how the excess energy redshifts away. We establish regions in the parameters space of the model that are ruled out by observations. \cref{VaryingBBN} explores varying some BBN parameters with one of the model's two input parameters, while keeping the other fixed. In \cref{sEMPRESS}, we discuss the recent anomalous measurement of primordial $^4$He, and how EDE could alleviate this tension. Finally, we summarize our conclusions in  
\cref{Conclusion}.

\section{Cosmological Equations for $\Lambda$CDM}
\label{cosmological eq}
The standard $\Lambda$CDM model of cosmology is characterized by an expanding, flat, homogeneous and isotropic Universe which is described by a Friedmann-Lema\^itre-Robertson-Walker (FLRW) metric. The conservation equation given an FLRW metric is
\begin{equation}
    D_\mu T^{\mu}_{\ \nu}=\dot{\rho}+3\left( \frac{\dot{a}}{a} \right)(\rho+p)=0,
\label{conservation}
\end{equation}
where $\rho$ is the energy density, $P$ is the pressure, and $a$ is the scale factor. Applying this stress-energy tensor and the FLRW metric to the Einstein field equations results in the Friedmann equation
\begin{equation}
    H^2\equiv \left(\frac{\dot{a}}{a}\right)^2 =\frac{8\pi}{3m_{\rm pl}^2}\rho,
\end{equation}
where $H$ is the Hubble parameter and $m_{\rm pl}=1.22\times10^{19}~\rm GeV$ is the Planck mass. The conservation equation \cref{conservation}, can be applied to each stress-energy component (radiation, matter, dark energy, etc.), where each component exerts a different pressure according to its equation of state, resulting in a different scaling of energy with the scale factor. The terms in the equation can be rearranged to give
\begin{equation}
    a^{-3}\frac{\partial [\rho a^3]}{\partial t}=-3\frac{\dot{a}}{a}P.
\end{equation}

From this relation one can infer that non-relativistic matter scales as $\rho_m\propto a^{-3}$, relativistic matter (radiation) scales as $\rho_r\propto a^{-4}$, and dark energy (or a cosmological constant) is constant, $\rho_{\Lambda}\propto a^0$.
Since each energy component scales differently with $a$, the dominant component is different at each epoch through the cosmological evolution of the Universe. In particular, during the period of BBN, the Universe was radiation dominated in $\Lambda$CDM. In this work, we propose an EDE component during BBN.

In standard BBN (SBBN), we assume $\Lambda$CDM cosmology and the standard model of particle physics with the number of neutrino species fixed at $N_\nu=3$. The only parameter in the theory is the baryon-to-photon ratio $\eta_b$, which is determined to better than $1\%$ accuracy from independent CMB measurements at later times~\cite{Planck,Cyburt:2015mya}.
To simplify the treatment here, we assume an instantaneous neutrino decoupling somewhat before the point when electron-positron annihilation happens (when the temperature of the primordial plasma is $T\sim 0.5$ MeV). BBN is also sensitive to the neutron lifetime $\tau_n$, nuclear reaction rates, and the effective relativistic degrees of freedom $N_{\rm eff}$
\begin{equation}
    N_{\rm eff}\equiv \frac{8}{7}\left(\frac{11}{4}\right)^{4/3}\frac{\rho_{\rm rad}-\rho_\gamma}{\rho_\gamma},
\end{equation}
where $\rho_{\rm rad}$ is the total radiation energy density 
\begin{align}
    \rho_{\rm rad}=&\ \rho_\gamma+N_\nu \rho_\nu \nonumber \\
    &+(\rm other \ radiation\text{-}like \ components).
\end{align}
Those quantities are either very well measured or set by SM calculations (e.g. see Ref.~\cite{deSalas:2016ztq} for $N_{\rm eff}$, the average from the particle data group for $\tau_n$~\cite{PDG2020}, and \cite{Serpico,Nag06} for nuclear reaction rates). 

The abundances of elements during BBN are governed by the Boltzmann equation in an expanding homogeneous and isotropic Universe
\begin{equation}
    \frac{\partial f}{\partial t}-Hp\frac{\partial f}{\partial p}=C[f],
\end{equation}
where $f$ is the distribution function of some species and $C[f]$ is the collision term which depends on the nuclear reaction rates. Assuming kinetic equilibrium with Maxwell Boltzmann distributions and ignoring Pauli blocking and Bose enhancement factors (see~\cite{Escudero:2018mvt,Dodelson:2003ft} for why those assumptions are justified), the Boltzmann equation for a reaction of the form $1+2\rightarrow 3+4$ can be simplified to 
\begin{equation}
a^{-3}\frac{d(n_1a^3)}{dt}= n_1^{(0)}n_2^{(0)}\langle \sigma v \rangle \left[ \frac{n_3n_4}{n_3^{(0)}n_4^{(0)}}-\frac{n_1n_2}{n_1^{(0)}n_2^{(0)}}  \right],
\label{Boltzmann}
\end{equation}
where $\langle \sigma v \rangle$ is the thermally averaged cross section, and $n_s$ is the number density of some species given by
\begin{equation}
    n_s=g_s\int \frac{d^3p}{(2\pi)^3}f,
\end{equation}
with $g_s$ being the spin degrees of freedom. The equilibrium number density $n_s^{(0)}$ for $m_s\gg T$ (Maxwell-Boltzmann) is 
\begin{equation}
    n_s^{(0)}=g_s\left(\frac{m_s T}{2\pi}\right)^{3/2}e^{-m_s/T}.
\end{equation}

Equation~(\ref{Boltzmann}) can be used for each reaction considered in BBN to build the nuclear reaction network to solve for elemental abundances. In building the nuclear reaction network (\cref{SectionBBN}), we consider a simple network of only the light elements (p, n, D, $^3$H, $^3$He, and $^4$He). Since we are using BBN as a probe of new physics, we avoid Be and Li due to the lithium problem where predictions exceed observations by a factor of 2--3 \cite{Ryan:1999rxn,Olive:1999ij}, although recent re-observations of halo stars suggest possible lithium depletion which may relax this tension \cite{Fields:2022mpw}. 

\section{The time-temperature relation}
\label{Tt}
The right-hand side of the Boltzmann equation has nuclear reaction cross sections $\langle \sigma v \rangle$ which are most easily expressed as functions of the temperature. So in order to solve the Boltzmann equation to obtain the abundances, one needs to find a time-temperature relation that describes the thermal evolution of the Universe given its contents.

The temperature evolution in time is obtained by considering conservation of total entropy
\begin{equation}
    \frac{dS}{dt}=0,
    \label{Sconservation}
\end{equation}
where the total entropy is $S=sV\propto sa^3$. The entropy density is defined as 
\begin{equation}
    s\equiv \frac{\rho+P}{T},
    \label{entropy}
\end{equation}
where we have neglected chemical potentials. For relativistic species, $P=\rho/3$ and so
\begin{equation}
  s= s_{\rm rad}=\frac{4}{3}\frac{\rho_{\rm rad}}{T}.
\end{equation}
After the neutrinos have decoupled from the electromagnetic plasma, the entropy of each sector is separately conserved. The entropy density of the Universe is 
\begin{equation}
    s=\frac{2\pi^2}{45}g_{* s}T_\gamma^3,
\end{equation}
where we define the effective relativistic entropy degrees of freedom
\begin{equation}
    g_{* s}\equiv \sum_B \left(\frac{T_a}{T_\gamma}\right)^3+\frac{7}{8}\sum_F\left(\frac{T_a}{T_\gamma}\right)^3,
\end{equation}
where $T_a$ is the temperature of a given species, the first sum is over bosonic species and the second sum is over fermionic species. Key quantities we want to keep track of are the photon temperature $T_\gamma$ and neutrino temperature $T_\nu$. For temperatures above about $3~{\rm MeV}$, weak interactions keep the neutrinos coupled to the photons and their temperatures are the same. After this point, the neutrinos decouple and the two sectors evolve separately with the entropy in each individually conserved. In the photon sector, the relevant degrees of freedom for the entropy are photons, electrons, and positrons and the effective relativistic entropy degrees of can be written as
\begin{equation}
    g_{* s\gamma}=2\left(1+\frac{s_{e^-}+s_{e^+}}{s_\gamma}\right),
\end{equation}
where $s_{e^\pm}$ is the entropy density of the electrons or positrons (computed using Fermi-Dirac statistics with the same temperature as the photons) and $s_\gamma$ that of the photons. As the Universe cools, $s_e$ decreases drastically, reheating the photons. In the neutrino sector, the effective relativistic entropy degrees of freedom does not change during this epoch. Plugging the entropy expressions for the two separate sectors into \cref{Sconservation} gives a system of coupled equations,
\begin{align}
\frac{dT_\gamma}{dt}=&-HT_\gamma\bigg(1+\frac{T_\gamma}{3g_{* s\gamma}}\frac{dg_{* s\gamma}}{dT_\gamma}\bigg)^{-1},
\label{TEvoG}
\\
\frac{dT_\nu}{dt}=&-HT_\nu.
\label{TEvoN}
\end{align}
The Hubble parameter in those equations is given by the Friedmann equation
\begin{equation}
    H=\sqrt{\frac{8\pi}{3m_{\rm Pl}^2}\rho_{\rm tot}},
\end{equation}
where in SBBN, the energy density
\begin{equation}
    \rho_{\rm tot}=\rho_{\gamma}(T_\gamma)+\rho_{e^-}(T_\gamma)+\rho_{e^+}(T_\gamma)+N_\nu\rho_\nu(T_\nu),
    \label{rho}
\end{equation}
is dominated by radiation.
Equations~(\ref{TEvoG}) and~(\ref{TEvoN}) can be solved numerically to obtain $T_\gamma(t)$ which is necessary to solve the BBN nuclear reaction network to obtain nuclear abundances. Finally, using the definition of the Hubble parameter, we have the additional equation
\begin{equation}
    \frac{da(t)}{dt}=a(t)H(t),
    \label{aEqn}
\end{equation}
which we also need to solve for when considering EDE scenarios to determine the excess energy's decay rate as a function of $a(t)$.

\section{The BBN nuclear reaction network}
\label{SectionBBN}
A general form of the Boltzmann equation for a reaction of the form $k+l\rightarrow i+j$, is  \cite{Wagoner,Esposito:2000hh}
\begin{equation}
\begin{aligned}
\dot{X}_i=\sum_{j,k,l}N_i\bigg(&\Gamma_{kl\rightarrow ij}\frac{X_l^{N_l}X_k^{N_k}}{N_l!N_k!} \\
&\quad\quad-\Gamma_{ij\rightarrow kl}\frac{X_i^{N_i}X_j^{N_j}}{N_i!N_j!}\bigg),
\label{general}
\end{aligned}
\end{equation}
where $N_s$ indicates how many times a species shows up in a reaction. This equation implicitly assumes that each nuclear species follows a Maxwell-Boltzmann distribution during nucleosynthesis, since their energies (essentially rest masses) are much larger than the temperature.

We consider here a simplified reaction network consisting of only six species (n, p, D,$^3$H, $^3$He, and $^4$He); therefore, we have a system of six coupled equations, one for each species.  For each reaction a species appears in, we write down a term as shown on the right-hand side of \cref{general}. The forward reaction rate $\Gamma_{kl\rightarrow ij}$ is defined as
\begin{equation}
\begin{aligned}
\Gamma_{kl\rightarrow ij}&\equiv \langle \sigma v \rangle_{kl\rightarrow ij} n_b  \\
& =  \eta_b\times \frac{2\zeta(3) }{\pi^2}T^3 \frac{f_{kl\rightarrow ij}}{N_A^{N_i-1}},
\end{aligned} 
\end{equation}
where $n_b$ is the baryon number density which could be expressed in terms of $\eta_b$ and the photon number density as in the second equality. We obtain the cross sections from reaction rate fits \cite{Serpico,Nag06}\footnote{We adopt reaction rate fits primarily from \cite{Serpico}, except for the reaction $d+n\rightarrow \gamma+ ^3\rm H$ which is adopted from \cite{Nag06}.} $f\equiv \langle \sigma v \rangle N_A^{N_i-1}$, with $N_A$ being Avogadro's number and $N_i$ the number of incoming nuclides (2 in all the cases we consider here).
The backward reaction rate $\Gamma_{ij\rightarrow kl}$ is defined as
\begin{equation}
\Gamma_{ij\rightarrow kl} \equiv \Gamma_{kl\rightarrow ij}R,
\end{equation}
where depending on the case, $R$ is calculated from 
\begin{enumerate}[(i)]
\item $k+l\rightarrow i+j$:
\begin{equation}
R\equiv\frac{n_i^{(0)}n_j^{(0)}}{n_k^{(0)}n_l^{(0)}}=\frac{X_{i,eq}X_{j,eq}}{X_{k,eq}X_{l,eq}},
\end{equation}
\item $k+\gamma\rightarrow i+ j$:
\begin{equation}
R\equiv\frac{n_i^{(0)}n_j^{(0)}}{n_k^{(0)}n_b}=\frac{X_{i,eq}X_{j,eq}}{X_{k,eq}}.
\end{equation}
\end{enumerate}
The equilibrium distribution for any species can be expressed in terms of only the neutron and proton abundances
\begin{equation}
\begin{aligned}
X_{i,eq}\equiv &\frac{n_i^{(0)}}{n_b}\simeq\frac{g_i}{2}\left(\zeta(3)\sqrt{\frac{8}{\pi}}\right)^{A_i-1}A_i^{3/2}  \\
&\times\left(\frac{T}{m_p}\right)^{\frac{3}{2}(A_i-1)}\eta_b^{A_i-1}X_p^{Z_i}X_n^{A_i-Z_i}e^{B_i/T},
\end{aligned}
\label{Xeq}
\end{equation}
where $A_i$, $Z_i$ and $B_i\equiv \sum_n m_n-m_i$ are the mass number, atomic number, and binding energy of the species respectively, and $m_n$ is the mass of comprising nucleons. Here, we assume that the masses in the prefactor are just multiples of $m_p$ (e.g. $m_{^4\rm He}\simeq4m_p$), but not in the exponent. In addition to determining reverse rates, the equilibrium abundance fractions determine the abundances at early times when the reaction rates are large enough for each species to obtain nuclear statistical equilibrium. 

We consider ten reactions that contribute the most to the BBN network of the light elements we are considering here (n, p, D,$^3$H, $^3$He, and $^4$He). The reactions and their corresponding contributing terms in the Boltzmann equation are shown in \cref{ReactionsTable}. Including those reactions in the Boltzmann equation \cref{general} results in the system of equations

\begin{align}
\dot{X}_n =& \dot{X}_{pn}-\dot{X}_{png}-\dot{X}_{ndd}-
\dot{X}_{ndt} + \dot{X}_{ptn} \nonumber \\
& - \dot{X}_{dng},
\\
\dot{X}_p =& -\dot{X}_{pn} -\dot{X}_{png} - \dot{X}_{dpg} -\dot{X}_{pdd} + \dot{X}_{he3dp} \nonumber \\
& - \dot{X}_{ptn} - \dot{X}_{tpg},
\\
\dot{X}_d=& \dot{X}_{png} - \dot{X}_{dpg} + 2\dot{X}_{ndd} + 2 \dot{X}_{pdd} + \dot{X}_{ndt} \nonumber \\
&- \dot{X}_{he3dp} - \dot{X}_{dng},
\\
\dot{X}_{^3 \rm H}=& -\dot{X}_{pdd} + \dot{X}_{ndt} - \dot{X}_{ptn} +\dot{X}_{dng} - \dot{X}_{tpg},
\\
\dot{X}_{^3 \rm He}=& \dot{X}_{dpg} - \dot{X}_{ndd} - \dot{X}_{he3dp} + \dot{X}_{ptn},
\\
\dot{X}_{^4 \rm He}=&-\dot{X}_{ndt} + \dot{X}_{he3dp} + \dot{X}_{tpg}.
\end{align}

\newcommand\Tstrut{\rule{0pt}{2.6ex}}    

\begin{table*}
\begin{tabular}{|c|c|}
\hline
Reaction & Contribution \\ 
\hline
\hline 
$n \leftrightarrow p$ \footnote{This includes the reactions $e^++n\leftrightarrow p+\overline{\nu}_e$, $p+e^-\leftrightarrow n+\nu_e$, and $n\leftrightarrow p+e^-+\overline{\nu}_e$.}& $\dot{X}_{pn}\equiv \omega_{pn}X_p-\omega_{np}X_n$ \Tstrut \\
$p+n \leftrightarrow \gamma +d$ & $\dot{X}_{png} \equiv \Gamma_{png}X_nX_p-\Gamma_{g pn}X_d$  \\
$d+p \leftrightarrow \gamma +\rm ^3He$ & $\dot{X}_{dpg}\equiv \Gamma_{dpg}X_dX_p-\Gamma_{g dp}X_{^3 \rm He}$ \\
$d+d \leftrightarrow n+\rm ^3He$ & $\dot{X}_{ndd}\equiv\Gamma_{ndd}X_nX_{^3 \rm He}-\Gamma_{ddn}\frac{X_d^2}{2}$ \\
$d+d \leftrightarrow p+\rm ^3H$ & $\dot{X}_{pdd}\equiv \Gamma_{pdd}X_pX_{^3 \rm H}-\Gamma_{ddp}\frac{X_d^2}{2}$ \\
$ {\rm ^3H}+d \leftrightarrow n+\rm ^4He $ & $\dot{X}_{ndt}\equiv \Gamma_{ndt}X_nX_{^4 \rm He}-\Gamma_{tdn}X_{^3 \rm H}X_d$ \\
${\rm ^3He}+d \leftrightarrow p+\rm ^4He $ & $\dot{X}_{he3dp}\equiv \Gamma_{he3dp}X_dX_{^3 \rm He}-\Gamma_{phe3d}X_pX_{^4 \rm He}$\\
${\rm ^3He}+n \leftrightarrow p+\rm ^3H$ & $\dot{X}_{ptn}\equiv \Gamma_{ptn}X_pX_{^3 \rm H}-\Gamma_{npt}X_nX_{^3 \rm He}$ \\
$d+n \leftrightarrow \gamma +\rm ^3H$ & $\dot{X}_{dng}\equiv \Gamma_{dng}X_dX_n-\Gamma_{gdn}X_{^3 \rm H}$ \\
${\rm ^3H}+p \leftrightarrow \gamma +\rm ^4He$ & $\dot{X}_{tpg}\equiv \Gamma_{tpg}X_{^3 \rm H}X_p-\Gamma_{gtp}X_{^4 \rm He}$ \\
\hline
\end{tabular}
\caption{The ten reactions considered in the BBN network along with their contributing terms.}
\label{ReactionsTable}
\end{table*}

The initial conditions for this set of equations come from \cref{Xeq}, and the constants needed to find equilibrium abundances (and consequently the reverse rates) are shown in \cref{constants}.

\begin{table}[t]
\begin{tabular}{|c|cccc|}
\hline
$N$ & $g$ & $A$ & $Z$ & $m\ (\rm m_u)$ \\
 \hline
 \hline
$n$		& 2 & 1 & 0 & 1.00866 \\
$p$		& 2 & 1 & 1 & 1.00728 \\
$d$		& 3 & 2 & 1 & 2.01410 \\
$^3$H	& 2 & 3 & 1 & 3.01605 \\
$^3$He 	& 2 & 3 & 2 & 3.01603 \\
$^4$He	& 1 & 4 & 2 & 4.00260 \\
\hline
\end{tabular}
\caption{Nuclear properties for nuclides used, where $g$ is the spin degrees of freedom, $A$ is the atomic mass number (number of protons + number of neutrons), $Z$ is the atomic number (number of protons), and $m$ is the isotope mass in atomic mass units. The listed masses are isotope masses \cite{nuclear}. Subtract $Zm_e$ to get the nuclear mass.}
\label{constants}
\end{table}

We fix BBN input parameters, namely the baryon-to-photon ratio $\eta_b\times 10^{10}\equiv \eta_{10}=6.104\pm0.058$, neutron lifetime $\tau_n=879.4\pm 0.6$ s, as reported by \cite{PDG2020,Planck,Fields:2019pfx}. The simplified treatment of temperature evolution does not take into account non-instantaneous neutrino decoupling and radiative effects which in the standard case would result in an $N_{\rm eff}=3.045$~\cite{deSalas:2016ztq}. To approximate this effect, we take $N_\nu=3.045$, which in the SBBN case gives the effective result of $N_{\rm eff}=3.045$ as desired. 

\begin{figure*}
\includegraphics[scale=0.5]{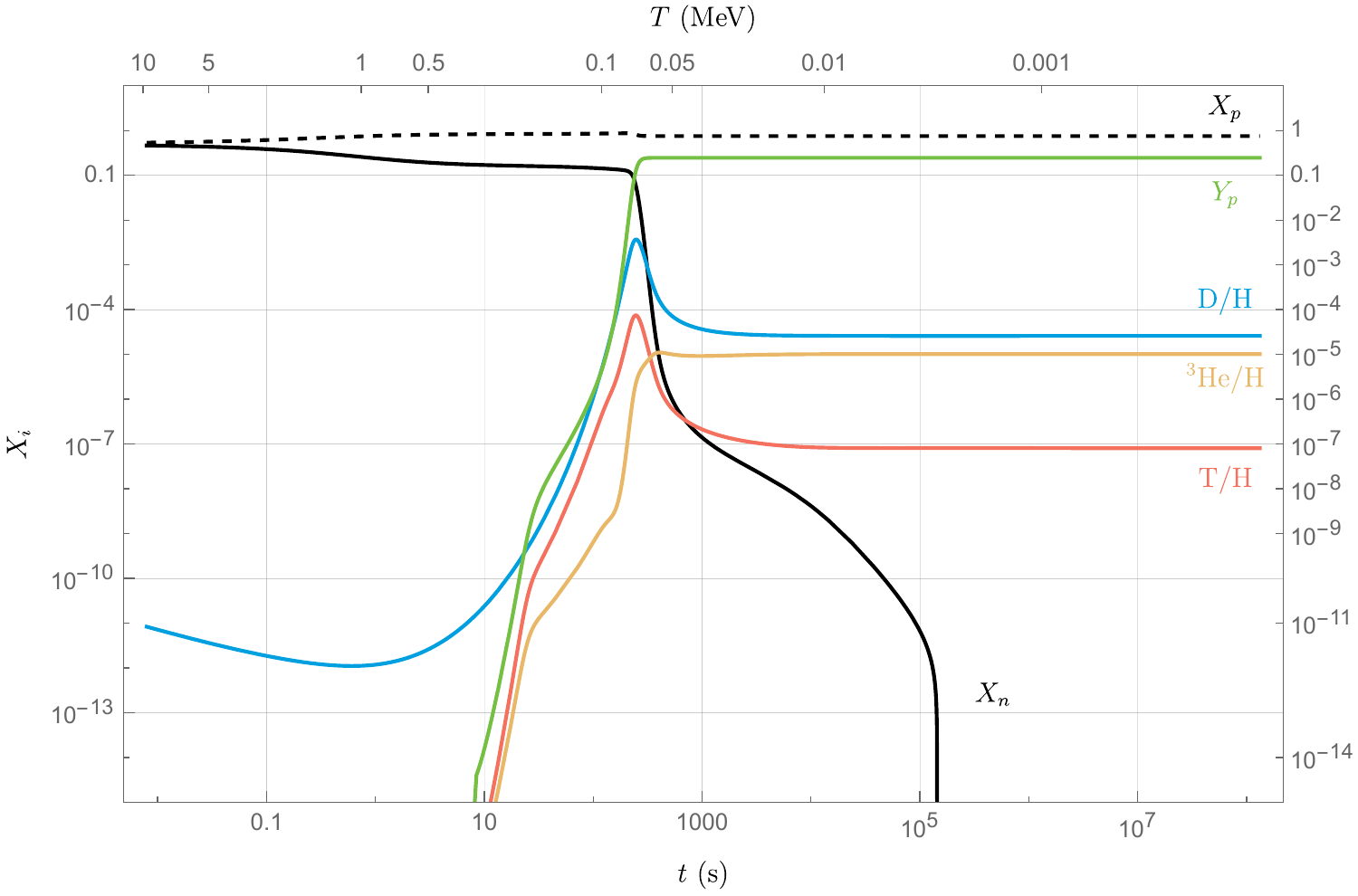}
\caption{The nuclear abundances that result from solving the system of equations. The abundances of D, T, and $^3$He are normalized to the hydrogen (proton) abundance, and $Y_p$ is the helium mass fraction.}
\label{Results}
\end{figure*}

Solving this system results in the standard picture for SBBN as shown in \cref{Results}, which shows the evolution of elemental abundances, consistent with previous more detailed BBN work in the literature \cite{Pospelov:2010hj,Cyburt:2015mya,Pitrou:2018cgg,Serpico}.

\section{Early Dark Energy During BBN}
\label{EDE_BBN}
The treatment in \cref{Tt,SectionBBN} so far has been entirely within standard $\Lambda$CDM and SBBN. We now modify the equations to introduce an EDE component. We consider a simple model with a constant dark energy present at early times which transitions into a different form at some critical time. In particular, in order to correctly describe the subsequent formation of the CMB, this new component must redshift like radiation or faster. Hence, this model we consider has only two parameters: the amount of constant EDE present $\rho_{\rm \Lambda}$ and the critical temperature at which the ``phase transition" happens $T_{\rm crit}$ where it transitions into a decaying component. These modifications directly change the time-temperature equations in \cref{Tt}, which consequently changes the outcomes of BBN, allowing us to put constraints on the two parameters.  

Most generically, the first addition to be made is to include an extra term in \cref{rho} to account for the extra EDE. Generally, this term should be constant up to a certain time; then it decays depending on which scenario we are considering.
The other significant physical quantity to consider is entropy. When EDE is still a constant, there is no additional entropy since the equation of state parameter for a constant energy density is $w=-1$, so its entropy is zero. After the transition, which we approximate as instantaneous, we consider three cases for the equation of state of this new component: (1) $w=1/3$, coupled to the SM plasma with associated entropy injection (which we dub ``transition into photons'' for short); (2) $w=1/3$, uncoupled to the SM plasma (which we dub ``transition into dark radiation'' for short); and (3) $w=1$, uncoupled to the SM plasma (``transition into kination''). 

\subsection{Transition into SM photons}

\begin{figure}
    \centering
    \includegraphics[scale=0.5]{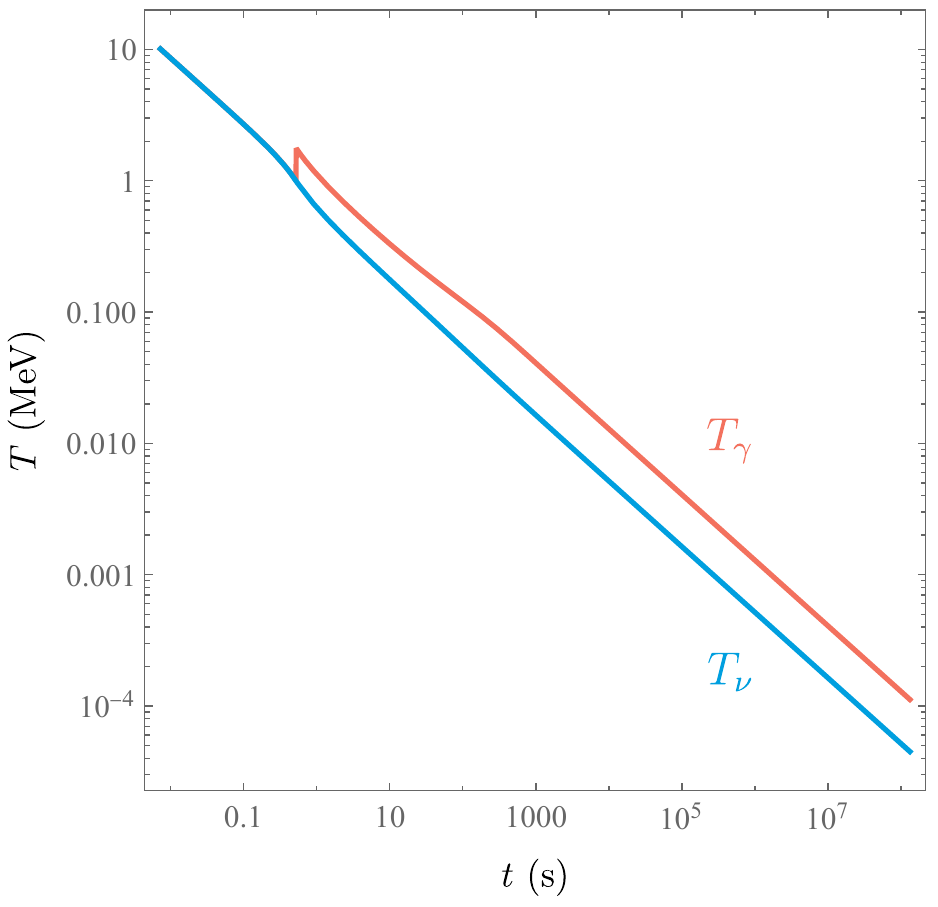}
    \caption{Photon and neutrino plasma temperature evolution in the presence of an EDE, $\rho_{\Lambda}^{1/4}=2$ MeV, that transitions to SM photons at $T_{\rm crit}=1$ MeV. The discontinuity in the photon temperature is a result of our simplifying assumption of an instantaneous transition.}
    \label{TEvoPlot}
\end{figure}

In this case, EDE transitions into regular SM photons which are in thermal contact with the rest of the plasma. The additional energy density takes the form 
\begin{equation}
    \rho_{\rm EDE}=
    \begin{cases}
\rho_\Lambda, \quad & t<t_{\rm crit}\\
0, \quad & t>t_{\rm crit}
    \end{cases},
    \label{rhoK}
\end{equation}
where $t_{\rm crit}$ is the critical time when the temperature is equal to $T_{\rm crit}$. Regardless of the scenario we are considering, we can determine $t_{\rm crit}$ by solving the Friedmann equation up to a sufficiently large time and then finding the time that satisfies $T(t_{\rm crit})=T_{\rm crit}$.
In this specific scenario, to preserve energy density conservation, our assumption of an instantaneous transition causes the photon temperature to suddenly increase at $t_{\rm crit}$. If we denote the temperature just before the transition as $T_-$, then the temperature just after the transition, $T_+$, can be found by setting the total energy density before and after the transition to be equal
\begin{equation}
\rho_\gamma(T_+)+2\rho_e(T_+)=\rho_\gamma(T_-)+2\rho_e(T_-)+\rho_\Lambda,
\end{equation}
which can be solved numerically. Therefore in this regard, the thermal evolution of the Universe is split into two steps, one before the critical point where the total energy density includes $\rho_\Lambda$ and the other after the critical point where the total energy density is that of standard $\Lambda$CDM but with the initial temperature given by $T_+$. The result of those numerical solutions of \cref{TEvoG,TEvoN} are the functions $T_\gamma(t)$ and $T_\nu(t)$, which are necessary in determining the outcomes of BBN. An example case for input parameters $\rho_{\Lambda}^{1/4}=2$ MeV and $T_{\rm crit}=1$ MeV is shown in \cref{TEvoPlot}, where the sharp increase in $T_\gamma$ corresponds to the critical point where EDE transitions to SM photons. This choice of parameters is ruled out by BBN, but we pick those large numbers to demonstrate the effect EDE would have. 
Importantly, we note that we do not choose parameters that lead to the $T_\gamma>3~\rm MeV$ after the transition since, in such a case, the neutrinos would recouple to the rest of the plasma and BBN would proceed as in the standard case.\footnote{Note that this is conservative compared to the more stringent condition of $T_\gamma\gtrsim 4~\rm MeV$ for standard BBN as obtained in Ref.~\cite{Hasegawa:2019jsa} when reheating into purely hadronic SM states.}

\begin{figure}
\centering
\includegraphics[scale=0.5]{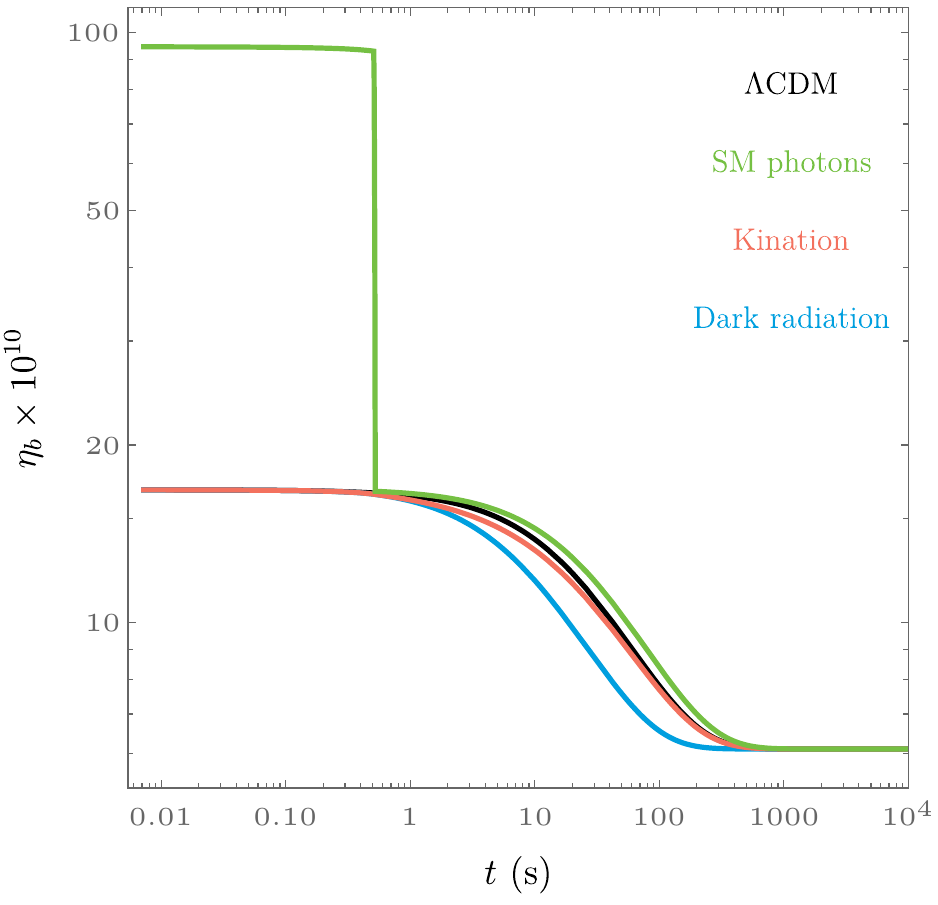}
\caption{The change in the baryon-to-photon ratio $\eta_b$ with time in each of the scenarios considered here with example input parameters  $T_{\rm crit}=1$ MeV and $\rho_{\Lambda}^{1/4}=2$ MeV. The smooth transition is due to electron-positron annihilation.}
\label{etabComparison}
\end{figure}

In addition to a time-temperature relation, since this scenario results in an increased number of photons, the baryon-to-photon ratio $\eta_b$ is expected to decrease after the critical point. Since the baryon number density is proportional to $a^{-3}$, and the photon number density is proportional to $T_\gamma^3$, then the baryon-to-photon ratio changes according to
\begin{equation}
\begin{aligned}
    \frac{\eta_b}{\eta_b|_{\rm initial}} &=\left(\frac{aT_\gamma|_{\rm initial}}{aT_\gamma}\right)^3
    \\ &=\left(\frac{T_\gamma/T_\nu|_{\rm initial}}{T_\gamma/T_\nu}\right)=\left(\frac{T_\nu}{T_\gamma}\right)^3,
\end{aligned}
\end{equation}
where in the second equality we used the fact that the neutrino temperature is approximately proportional to $a^{-1}$ (minimal neutrino interactions) and the last equality comes from the fact that $T_\gamma=T_\nu$ at early times. The ratio changes when electrons and positrons annihilate (since the photon number density increases) and again once the EDE gets converted into SM photons (see also Ref.~\cite{Sobotka:2022vrr} for an example of this). An example of this behaviour is shown in the green curve in \cref{etabComparison} for the example case with parameters $\rho_{\Lambda}^{1/4}=2$ MeV and $T_{\rm crit}=1$ MeV as before. The initial ratio is calculated such that we obtain a final baryon-to-photon ratio $\eta_b=6.104\times 10^{-10}$,
as observed by Planck 2018~\cite{Planck,Fields:2019pfx}.

\begin{figure*}
  \centering
  \subfigure[]
  {\includegraphics[scale=0.5]{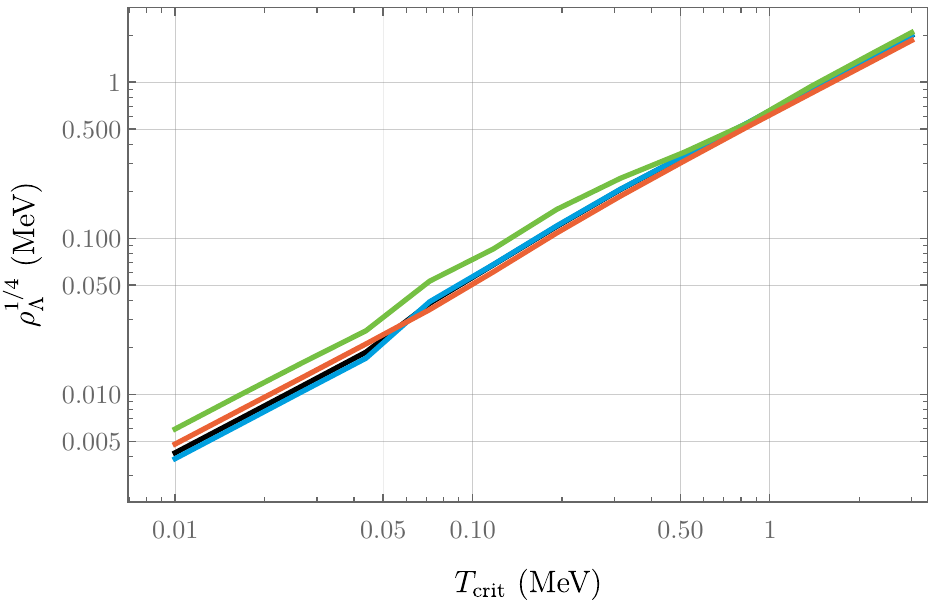}
  \label{rhoGTcrit}}
  \quad
  \subfigure[]
  {\includegraphics[scale=0.5]{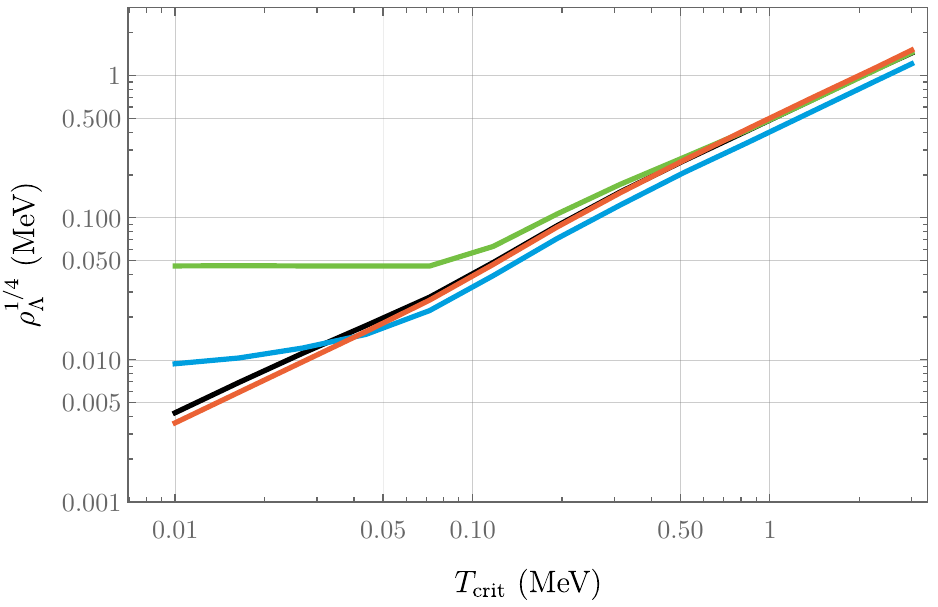}
  \label{rhoGTcritDR}}
  \quad 
   \subfigure[]
   {\includegraphics[scale=0.5]{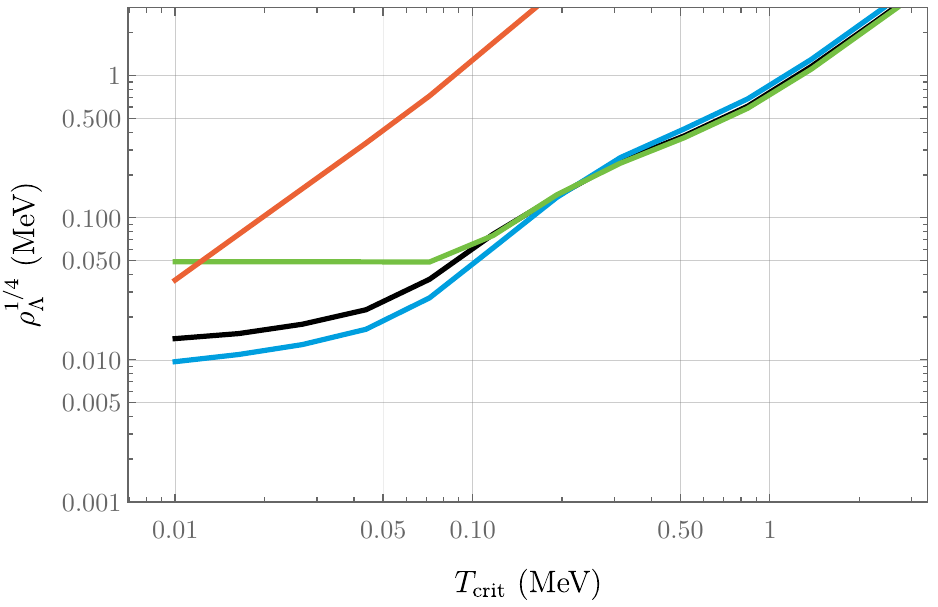}
   \label{rhoGTcritK}}
   \caption{Limits on the parameter space of $\rho_\Lambda$ and $T_{\rm crit}$ from D/H abundance (blue curve), $Y_p$ (green curve), and $N_{\rm eff}$ (red curve). The combined limit from all three observables is shown in the thick black curve. The region above the curves is excluded by this model. (a) is the case of a decay of EDE into SM photons, (b) is a decay into dark radiation, and (c) is a decay into kination. }
\end{figure*}

Each pair of points in the parameter space of $\rho_\Lambda$ and $T_{\rm crit}$ results in a different outcome for the abundances of elements at the end of BBN. Here we consider constraints from D~\cite{Cooke:2013cba,*Riemer-Sorensen:2014aoa,*Balashev:2015hoe,*Riemer-Sorensen:2017pey,*2018MNRAS.477.5536Z,*Cooke:2017cwo} and $^4$He~\cite{Aver:2020fon,*Valerdi:2019beb,*Fernandez:2019hds,*Kurichin:2021ppm,*2020ApJ...896...77H,*2021MNRAS.505.3624V,*2022MNRAS.510..373A} abundances as averaged by the PDG~\cite{PDG2020}:
\begin{align}
    \rm D/H\times 10^5 &= 2.547\pm 0.025, \\
    Y_p &= 0.245\pm 0.003,
    \label{eq:Ypmeas}
\end{align}
 in addition to constraints from $N_{\rm eff}$ as measured by Planck~\cite{Planck}:
\begin{align}
        N_{\rm eff} &= 2.99\pm 0.17.
\end{align}
We establish exclusion regions that result in abundances and $N_{\rm eff}$ within 1$\sigma$ from the observed values, taking into account theoretical uncertainties on the abundance coming from uncertainties on the input parameters $\eta_b$ and $\tau_n$ as well as the nuclear uncertainties as parameterized in~\cite{Serpico}. 
Those regions for the two input parameters in this scenario are shown in \cref{rhoGTcrit}, where the EDE transitions into SM photons.

The yellow curve indicates the limit where $\chi^2<1$ for the D/H abundance (D abundance normalized by H abundance), where 
\begin{equation}
    \chi^2=\frac{\left(\rm D/H_{calculated}-D/H_{observed}\right)^2}{\sigma_{\rm obs}^2+\sigma_{\rm theoretical}^2},
\end{equation}
for one degree of freedom. The region above the curve is ruled out. Similarly, the green curve is the $\chi^2<1$ limit for helium abundance $Y_p$, and the red curve is the limit for $N_{\rm eff}$. The thick black curve is the combined limit given all three observations (where $\chi^2<3.506$ for three degrees of freedom). The axes are cutoff at 3 MeV because we do not consider transitions that recouple the neutrinos since, as mentioned above, that results in standard BBN.

At any given time after the critical point, this scenario results in a temperature higher than the corresponding temperature at the same time in the case of a $\Lambda$CDM cosmology as shown in \cref{TComparison} (green curve) with example input parameters as before. This is effectively equivalent to a Universe that has expanded at a slower rate than $\Lambda$CDM, and so results in lower final abundances since fewer neutrons were available at the time of nucleosynthesis. The lower abundance of neutrons available at the start of BBN is caused by more time elapsing in this scenario, thus allowing more neutrons to convert to protons via free neutron decay. In addition, $N_{\rm eff}$ is lowered here since the photon energy density has increased. 

The analysis in this section could also be applied for a case of negative initial EDE density $\rho_{\Lambda}<0$. Doing so, we find similar limits on $\left|\rho_{\Lambda}\right|$. We limit this work to positive EDE. For a discussion of the implications of a negative EDE, see \cite{Pospelov:2010hj,Takahashi:2022cpn}.

\subsection{Transition into dark radiation}

\begin{figure}
    \centering
    \includegraphics[scale=0.5]{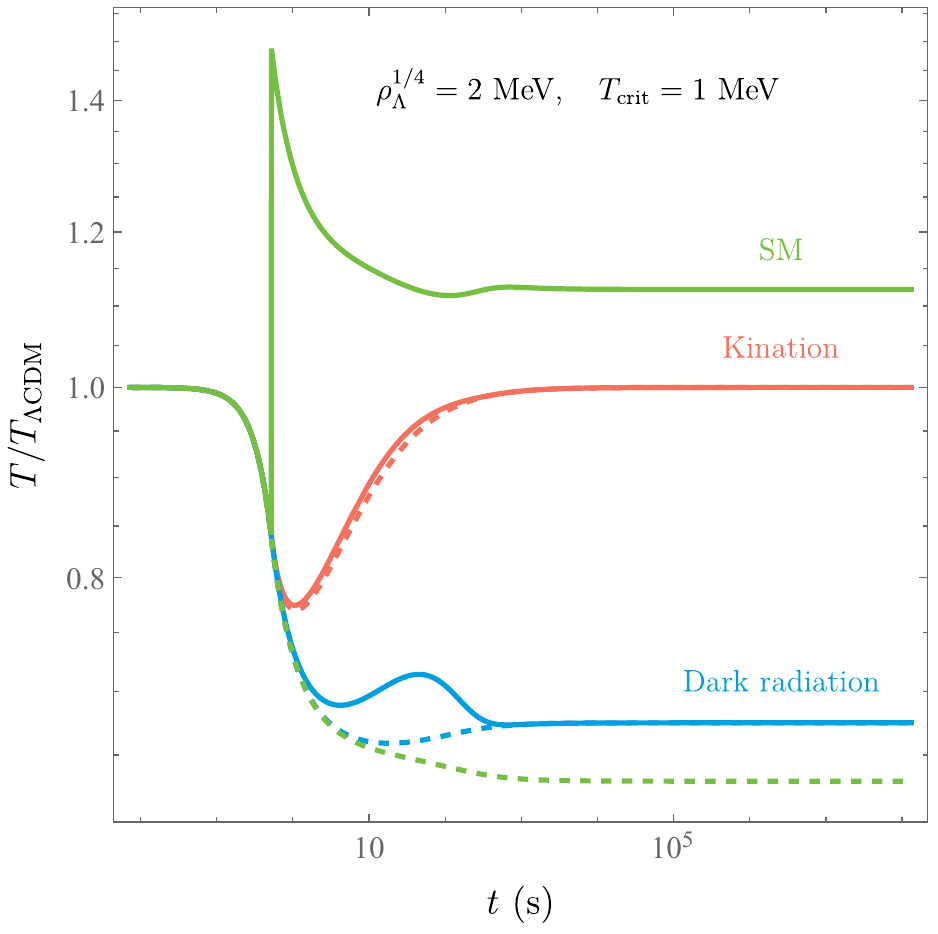}
    \caption{Photon (solid lines) and neutrino (dashed lines) temperature ratios of a given scenario relative to $\Lambda$CDM in the three scenarios we consider.}
    \label{TComparison}
\end{figure}

At the transition point in this case, we assume that the EDE instantaneously becomes ``dark radiation". In other words, it starts to redshift $\propto a^{-4}$ and does not interact with the SM plasma (no entropy injection).
Since neutrinos also redshift as $\propto a^{-4}$ and do not interact with the plasma, this scenario is roughly equivalent to an EDE transition into neutrinos. The additional energy density takes the form 
\begin{equation}
    \rho_{\rm EDE}=
    \begin{cases}
\rho_\Lambda, \quad & t<t_{\rm crit}\\
\rho_\Lambda\left(\frac{a_{\rm crit}}{a(t)}\right)^{3(1+w)}, \quad & t>t_{\rm crit}
    \end{cases},
    \label{rhoK}
\end{equation}
where $a_{\rm crit}$ is the value of the scale factor at the critical time and $w$ is the equation of state parameter which is equal to $1/3$ in this case. Unlike the previous case, there is no jump in the photon temperature or $\eta_b$ at the critical point since there is no entropy injection ($T_-=T_+$). Owing to the added energy density, this scenario results in a faster expansion of the Universe and thus a lower temperature of the plasma at any given time when compared to $\Lambda$CDM as can be seen in \cref{TComparison} (blue curve). The small bump is due to the time shift of electron-positron annihilation which alters the photon temperature (solid) relative to the neutrino temperature (dashed). As in the previous case, the baryon-to-photon ratio also changes with time; however, it does not exhibit the jump (blue curve for this case, versus green curve for the previous case in \cref{etabComparison}), since no entropy is injected and so the ratio is not changed due to EDE. The smooth change due to electron-positron annihilation however is still present. Since the expansion is faster in this scenario, annihilation occurs earlier than $\Lambda$CDM (black curve), and so the change in $\eta_b$ shifts earlier as shown in \cref{etabComparison}. 

The limits on the parameter space in this scenario are shown in \cref{rhoGTcritDR}. Since no entropy is injected into the plasma, and consequently $\eta_b$ does not change, the limit on $\rho_{\Lambda}$ from elemental abundances plateaus at late-time transitions (low $T_{\rm crit}$). This is because when the transition occurs at a critical temperature well below $T_{\rm bn}$, BBN processes have already largely finished. Therefore, the maximum allowed EDE does not depend on the critical point, and is rather just an upper limit on the allowed excess energy during BBN. This shows up in \cref{rhoGTcritDR} as the plateau at low $T_{\rm crit}$ (especially visible for limits from $^4$He since its synthesis finished relatively quickly, compared to D which exhibits late-time depletion). This is in contrast to the previous case where a plateau did not occur since there is also entropy injection in that scenario. The direct result of entropy injection is to force the initial value of $\eta_b$ to change (see SM photon case in \cref{etabComparison}) so that its late-time value agrees with CMB measurements. However, the value of $\eta_b$ during BBN does affect the outcome of elemental abundances. 
Limits from $N_{\rm eff}$ do not plateau (in either scenario) since they depend on the much later measurement during the CMB epoch, and the amount of allowed extra radiation continues to be related to the time it is injected.

Unlike the previous case, the abundances of D and $^4$He in this scenario are higher than in SBBN. This is because more neutrons are available at the time of nucleosynthesis since the Universe expanded more quickly leaving less time for them to get converted into protons (or decay). More available neutrons in the reaction network result in higher abundances for D and $^4$He (among others). In addition, since the extra added energy transitions into (dark) radiation, this scenario results in an increased $N_{\rm eff}$.

\subsection{Transition into kination}

The last scenario we consider is EDE that transitions into a component that redshifts $\propto a^{-6}$, called kination~\cite{Joyce:1996cp}. The energy density here is characterized by a scalar field whose dynamics are mainly dominated by its kinetic energy. A scalar field can be treated as a perfect fluid; thus, its equation of state is given by 
\begin{equation}
w=\frac{\rm KE-PE}{\rm KE+PE}\simeq 1.
\end{equation}
This is the maximum equation of state allowed by causality, i.e. the speed of sound is equal to the speed of light, and has the fastest possible redshifting of energy density ($\rho\propto a^{-6}$). Therefore a kination component of the energy density becomes subdominant quickly compared to other components (e.g. radiation, matter, etc.), or it leads to a slowed expansion of the Universe if it is the dominant component present.
A kination period could be induced by a fast-rolling inflaton field at the end of inflation where the field has a steep potential~\cite{Gouttenoire:2021jhk,Spokoiny:1993kt}, analogous to the brief period where the energy density is dominated (or nearly dominated) by $\rho_{\Lambda}$ in our treatment here. 

Since the equation of state parameter is set to $1$, the EDE density is given by
\begin{equation}
    \rho_{\rm EDE}=
    \begin{cases}
\rho_\Lambda, \quad & t<t_{\rm crit}\\
\rho_{\Lambda}\left(\frac{a_{\rm crit}}{a(t)}\right)^{6}, \quad & t>t_{\rm crit}
    \end{cases}.
    \label{rhoEDE}
\end{equation}
The extra energy density redshifts very quickly after the transition, bringing the temperature of the plasma approximately back to its $\Lambda$CDM value at late times as shown in the red curve in \cref{TComparison}. This scenario results in a faster expansion of the Universe (similar to the previous case), and so the change in $\eta_b$ due to annihilation is also shifted earlier as shown in the green curve in \cref{etabComparison}. The shift is less dramatic as in the previous scenario since expansion is not as fast due to the quick disappearance of the EDE. 

The limits on the parameter space in this scenario are shown in \cref{rhoGTcritK}. As in the case of dark radiation, no entropy is injected here, and so the limits from elemental abundances plateau at low critical temperatures, whereas limits from $N_{\rm eff}$ do not. The limits from $N_{\rm eff}$ are much weaker here compared to other scenarios since kination results in only a small change of the expansion rate of the Universe.

\section{Varying other BBN parameters}
\label{VaryingBBN}
In the previous section, we put constraints on the input parameters $\rho_\Lambda$ and $T_{\rm crit}$, given constant SBBN parameters
\begin{align}
    \tau_n &=879.4\pm0.6\ \rm s, \\
    \eta_b &=6.104\pm0.058\times 10^{-10}, \\
    N_\nu &= 3.045.
\end{align}
We consider here the dependence of the EDE input parameters on each of those for all the scenarios considered in the previous section. 

\begin{figure*}
\includegraphics[scale=0.5]{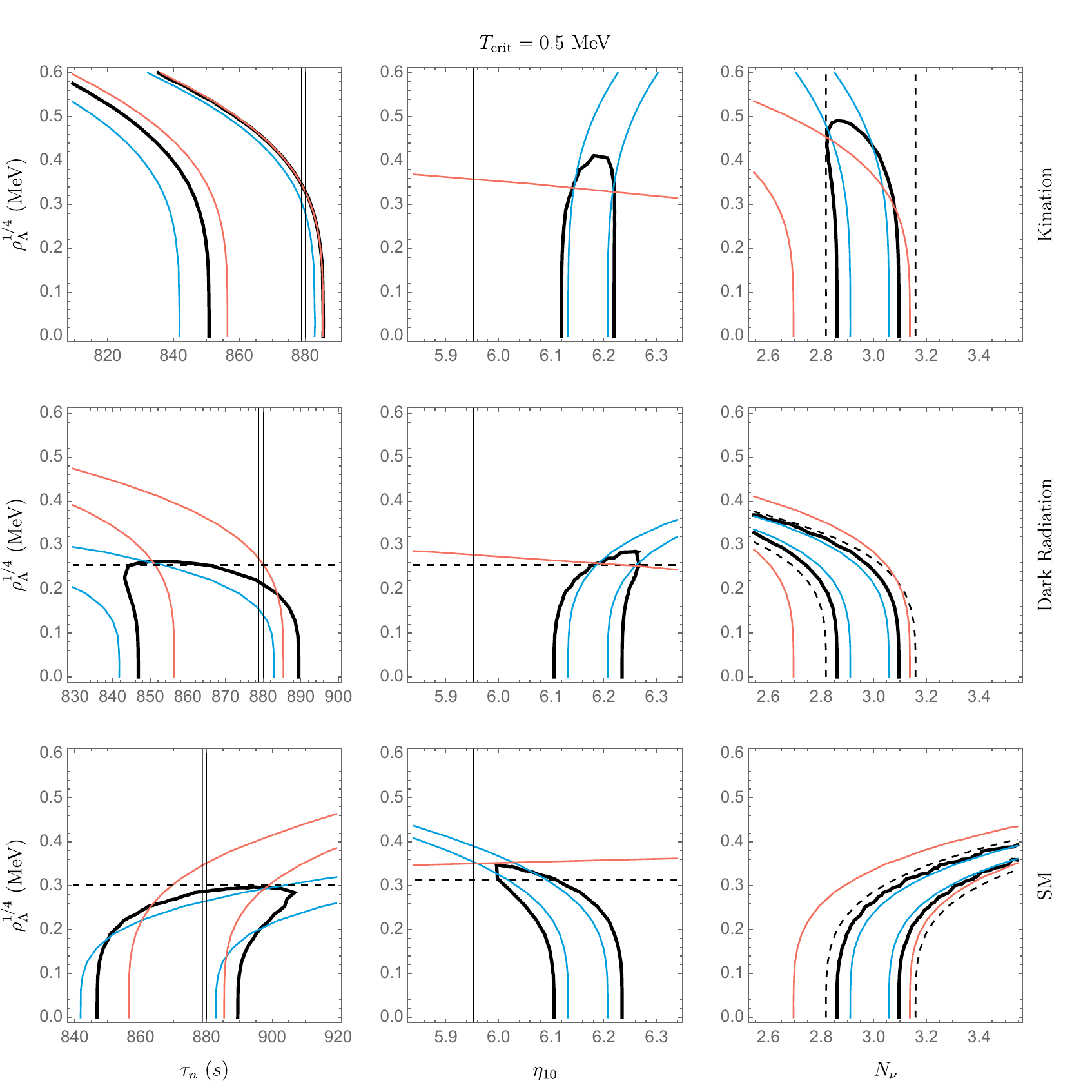}
\caption{Dependence of $\rho_\Lambda$ on BBN parameters with a fixed $T_{\rm crit}=0.5$ MeV, for a transition of EDE into SM photons (bottom row), dark radiation (middle row), and kination (top row). Space inside the contours results in outcomes that are within $1\sigma$ from observed D/H (blue), $Y_p$ (red), and $N_{\rm eff}$ (dashed) \cite{PDG2020,Planck}. Thick contours are the combined $1\sigma$ limits, and thin vertical lines are the $1\sigma$ errors on the observed $\tau_n$ and $\eta_{10}$.}
\label{GridF}
\end{figure*}
To explore the dependence of $\rho_{\Lambda}$ on the BBN parameters in a 2D parameter space, we set $T_{\rm crit}=0.5$ MeV. Since the effect of adding extra energy ultimately results in altering the temperature evolution, we expect a correlation between the amount of EDE and the neutron lifetime. This is indeed visible in the first column of panels in \cref{GridF}. The region inside the blue and red contours results in D/H and $Y_p$ (respectively) within $1\sigma$ of the observed values. The dashed line (or contour) is the limit from $N_{\rm eff}$, where regions above the line are excluded. It is independent of $\tau_n$ as it is unrelated to nucleosynthesis. The thin gray lines indicate the measured value of neutron lifetime. The thick contour is the combined $1\sigma$ limit calculated by requiring $\chi^2<3.51$ (for three degrees of freedom), or $\chi^2<2.28$ (for two degrees of freedom) for cases where limits from $N_{\rm eff}$ cannot be applied. The direction of the correlation depends on whether the expansion is decreased (SM case, bottom row), or increased (dark radiation and kination cases, middle and top rows respectively). A decay of EDE into SM photons results in an increased plasma temperature compared to $\Lambda$CDM. This is equivalent to a relative decrease in the expansion of the Universe, and consequently more time elapsed when $T_{\rm bn}$ is achieved, leaving fewer neutrons available at nucleosynthesis. This effect is counteracted by an increased neutron lifetime, hence the positive correlation seen in the bottom left panel of \cref{GridF}. The reverse effect is visible for a decay of EDE into dark radiation or kination resulting in a negative correlation as shown in the middle and top left panels. The correlation is weaker for the case of kination since the expansion increase is not as strong as the case of dark radiation.   

By a similar argument, the baryon-to-photon ratio is correlated to the amount of EDE since nuclear abundances are sensitive to $\eta_b$ \cite{Schramm:1997vs,Fields:2019pfx,Cyburt:2015mya}. Since  $Y_p$  is logarithmically sensitive to $\eta_b$, the correlation is weak, as shown by the red lines with a small slope in the middle column of panels. An increase in either $\eta_b$ or $\rho_\Lambda$ for a decay of EDE into SM photons results in lowered D/H, giving a negative correlation as shown in the bottom panel of the middle column. For the other two cases, increasing $\rho_\Lambda$ increases D/H abundance, and so the correlation is positive (with kination being weaker as before).

Finally, the dependence of $\rho_\Lambda$ on the number of neutrinos $N_{\nu}$ is shown in the third column. Unlike the other two BBN parameters, $N_\nu$ does affect the thermodynamics and therefore changes the ratio $T_\gamma/T_\nu$, resulting in an altered $N_{\rm eff}$ (and thus limits on it). The limits are also present in the case of kination (top right panel), although the correlation is very weak.

\begin{figure*}[t]
\includegraphics[scale=0.5]{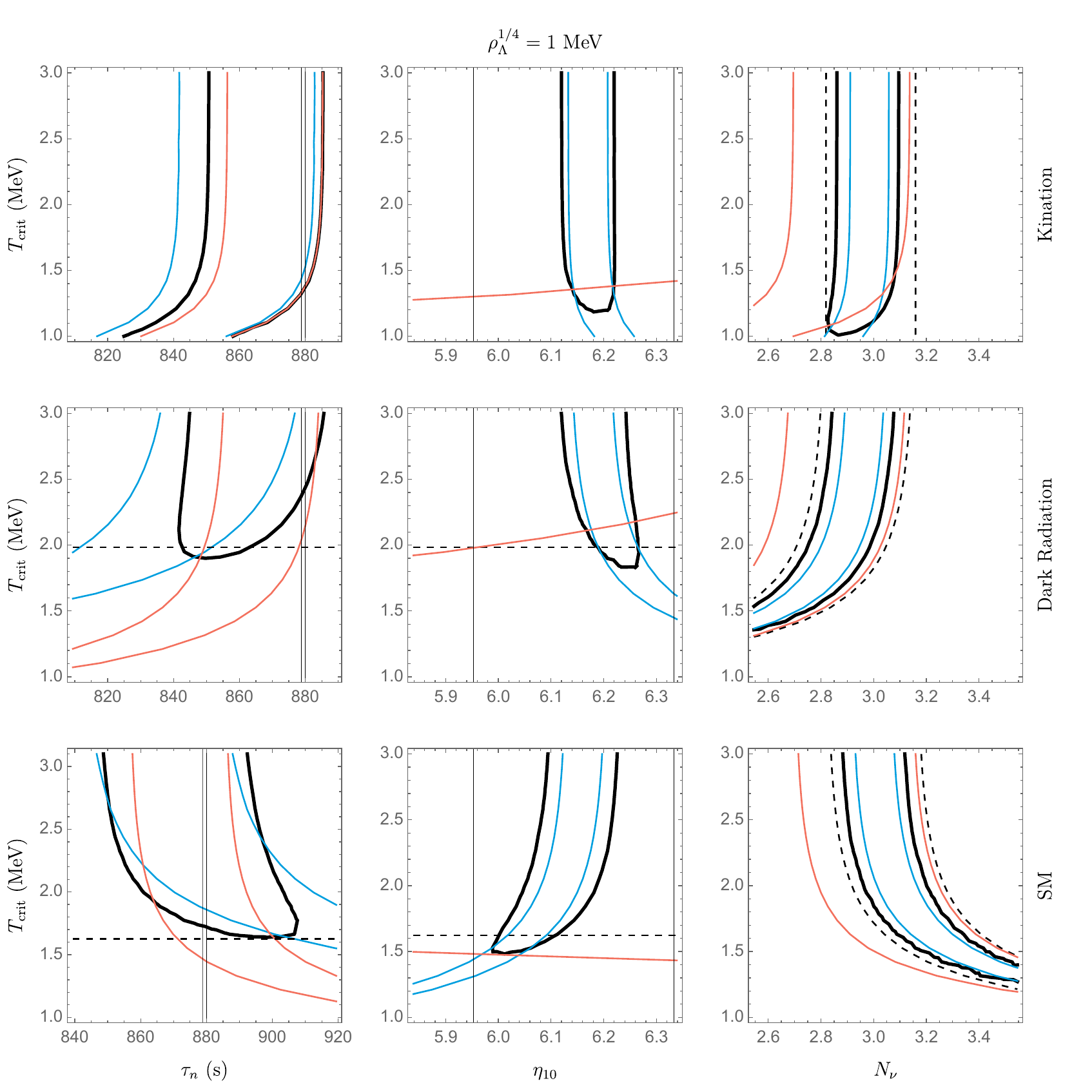}
\caption{Dependence of $T_{\rm crit}$ on BBN parameters for a fixed $\rho_\Lambda^{1/4}=1$ MeV. Layout and contours are the same as \cref{GridF}.}
\label{GridG}
\end{figure*}
 To see the dependence of $T_{\rm crit}$ on the BBN parameters in a 2D parameter space, we similarly choose $\rho_\Lambda^{1/4}=1$ MeV and vary the other parameters as shown in \cref{GridG}. The correlations follow a similar pattern as above but with the vertical axis flipped, where EDE's impact is weaker the larger $T_{\rm crit}$ is.

\section{Impact of a smaller $Y_p$}
\label{sEMPRESS}
Recently, the EMPRESS Collaboration has included new data in their determination of the primordial $^4$He abundance, resulting in $Y_{p}^{\rm EMPRESS}=0.237\pm0.003$~\cite{EMPRESS}. This is around $1.9\sigma$ below the value in Eq.~(\ref{eq:Ypmeas}) obtained from averaging the estimates in Ref.~\cite{Aver:2020fon,*Valerdi:2019beb,*Fernandez:2019hds,*Kurichin:2021ppm,*2020ApJ...896...77H,*2021MNRAS.505.3624V,*2022MNRAS.510..373A} that we use in our analysis above and $3.4\sigma$ below the value we predict with our SBBN code. While there is the possibility that this smaller value is the result of underestimated systematics or mismodeling, it has motivated the possibility that BBN was modified from the standard picture, e.g. varying the Higgs vacuum expectation value \cite{Burns:2024ods}, to potentially alleviate this tension. In this section we study whether EDE during BBN can explain a smaller $Y_p$ should the EMPRESS results turn out to be correct.

\begin{figure}[h]
    \centering
\includegraphics[scale=0.5]{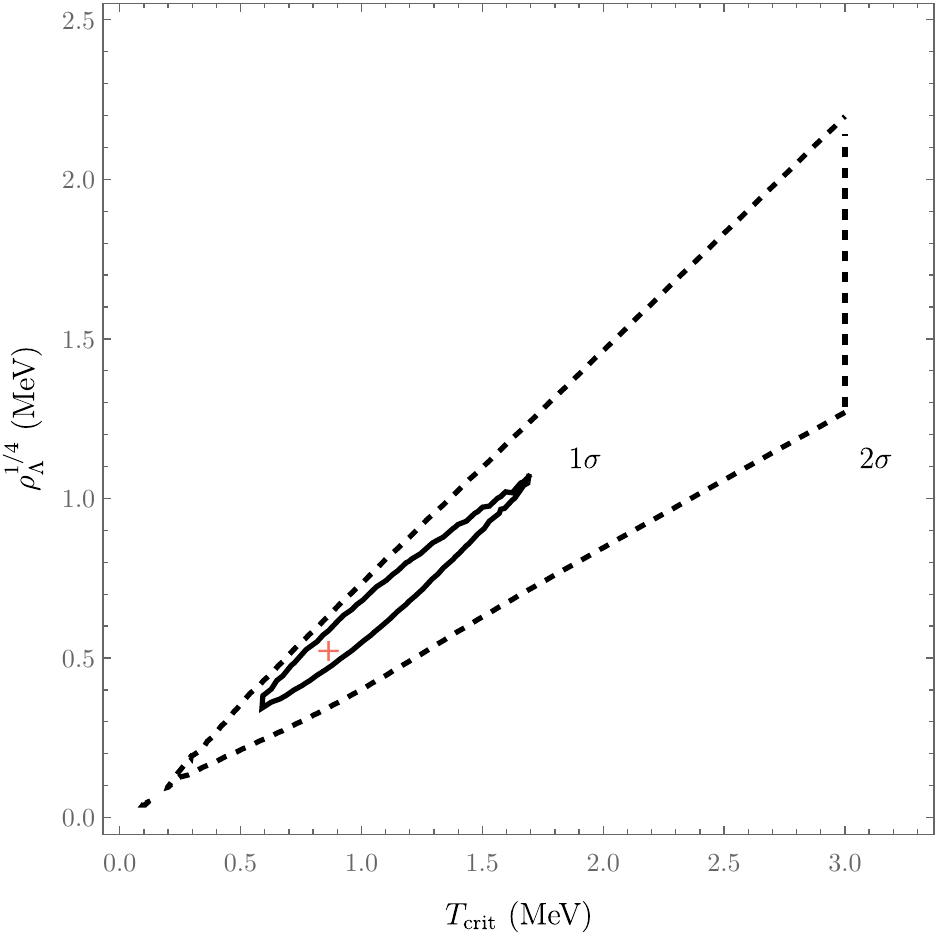}
    \caption{The best-fit regions (solid for $1\sigma$ and dashed for $2\sigma$) away from the best-fit point in the EDE model to relieve the primordial $^4$He abundance tension between EMPRESS and the PDG.}
    \label{deltaChi}
\end{figure}

Generically, the presence of EDE that transitions to uncoupled components increases $Y_p$, if SBBN parameters are held constant. However, allowing SBBN parameters like $N_{\rm eff}$ or $\eta_b$ to deviate from their standard values could permit a better fit to the new $Y_p$ value in the presence of EDE. Alternatively, one could consider a scenario with a negative EDE, which requires no deviation of SBBN parameters \cite{Takahashi:2022cpn}.    In light of the discussion in the previous section, we note here that EDE that transitions into photons coupled to the SM plasma with $T_{\rm crit}>T_{\rm bn}$ can reduce $Y_p$. We quantify this in Fig.~\ref{deltaChi}, showing the $1$ and $2\sigma$ limits in the $\rho_\Lambda^{1/4}$-$T_{\rm crit}$ plane which we determine by computing $\Delta \chi^2$ with respect to the best-fit point. The best-fit point at $\rho_{\Lambda}^{1/4}\simeq 0.540~\rm MeV$ and $T_{\rm crit}\simeq 0.864~\rm MeV$ is indicated with a red cross.
We note that this scenario leads to values of D/H and $N_{\rm eff}$ that are in worse agreement with the data than the SBBN predictions but provides an improvement in the overall fit by reducing $Y_p$. Also note that the hard cutoff of the $2\sigma$ contour comes from the fact that reheating the SM plasma above this temperature just reproduces SBBN.

\section{Conclusion}
\label{Conclusion}
The early history of the Universe is still largely a mystery. The earliest direct observation of the Universe comes from BBN. This picture is largely consistent with an expanding radiation-dominated Universe. However, the possibility of a non-negligible component of the Universe's energy density that redshifts like dark energy is motivated on several grounds: (i) if this happens around recombination, it can alleviate the Hubble tension, (ii) such a component of the Universe exists today in the form of dark energy and likely did so shortly after the big bang during inflation, and (iii) the effective field theory picture of vacuum energy indicates that its value changes during phase transitions generally.

In this work, we have studied the impact of an early dark energy component on the outcomes of BBN, the earliest direct probe of such a scenario that we have. Since primordial elemental abundances, as well as $N_{\rm eff}$, are well-measured quantities, we have used observations to put constraints on the parameters of this EDE model: the amount of EDE initially present before the transition $\rho_\Lambda$, and the temperature at which it transitions $T_{\rm crit}$. The model consists of an EDE that begins as a constant component of energy density (similar to the cosmological constant), and then it decays at $T_{\rm crit}$. 

We considered three different scenarios, namely a transition into SM photons, dark radiation, or kination. Each scenario impacts the thermal evolution of the Universe differently, resulting in different BBN outcomes and consequently different constraints on $\rho_\Lambda$ and $T_{\rm crit}$. We have also considered the relationships between the EDE input parameters $\rho_\Lambda$ and $T_{\rm crit}$ and the BBN input parameters $\tau_n$, $N_\nu$, and $\eta_b$ in the predicted light element abundances. We find correlations consistent with the general notion that in the case of a transition into SM photons, the Universe expands slower than in $\Lambda$CDM, whereas in the case of a transition into dark radiation or kination it expands faster. Lastly, we studied the possibility that recent hints of a smaller-than-expected primordial helium abundance can be explained in this setup, finding that a transition into photons before the deuterium bottleneck is able to improve the overall BBN fit in this case.

\begin{acknowledgments}
We thank John Coffey, Carlos de Lima, David Morrissey, Michael Shamma, and Tsung-Han Yeh for useful discussions and comments.
We also thank Adam Ritz for valuable feedback on the manuscript and helpful discussions. 
This work is supported in part by the Natural Sciences and Engineering Research Council of Canada (NSERC). TRIUMF receives federal funding via a contribution agreement with the National Research Council (NRC) of Canada.
\end{acknowledgments}


\bibliography{refs}

\end{document}